\documentclass[11pt,preprint]{aastex}

\begin{document}
\title{How to Nurture Scientific Discoveries Despite Their Unpredictable Nature}

\author{Abraham Loeb\\Institute for Theory \& Computation\\
Harvard University\\60 Garden St., Cambridge, MA 02138}

\begin{abstract}

The history of science reveals that major discoveries are not
predictable.  Naively, one might conclude therefore that it is not
possible to artificially cultivate an environment that promotes
discoveries.  I suggest instead that open research without a
programmatic agenda establishes a fertile ground for unexpected
breakthroughs.  Contrary to current practice, funding agencies should
allocate a small fraction of their funds to support research in
centers of excellence without programmatic reins tied to specific
goals.

\end{abstract}

\bigskip
\bigskip
\bigskip

{\it ~\\``As for the donkeys you lost three days ago, \\ 
do not worry about them...''}

\noindent
{\it 1 Samuel, Chapter 9, 20}

\section{Seeking Lost Donkeys but Finding a Kingdom}

The biblical story of Saul searching for lost donkeys and finding his
kingdom by chance, has an important moral for scientists. It is
essential not to define your research objectives too narrowly and open
your mind to discovering something completely different and more
exciting lurking at the periphery of your field of view.

Funding agencies are obligated to justify their use of taxpayers'
money over a period of several years. They are naturally driven to
fund low-risk research with predictable returns. Here I argue that to
maximize our long-term benefits, this approach has to change. In
particular, funding agencies should allocate a small fraction of their
funds (say 20\%) to open research in centers of excellence without
programmatic reins tied to specific goals. Such a funding scheme is
essential for promoting breakthroughs in the long run, since it
encourages researchers to take on high-risk projects with potentially
high gains but fundamentally unpredictable outcomes.

An example of an unexpected result is the discovery of the cosmic
microwave background by Arno Penzias and Bob Wilson, who were
attempting to reduce the noise in their state-of-the-art horn antenna
in 1965. They noticed a noise floor, which turned out to be the
radiation left over from the Big Bang. Interestingly, this watershed
discovery that forever changed our view of the Universe, was made at
Bell Labs and not at a premier research university.

It is common to think about short-term goals in funding physics, but
nurturing data-driven research with no programmatic goals promotes
innovation and brings unanticipated profits. The data component is
essential since extended periods of time without data allow
unrestrained growth of speculative theory bubbles which might have no
real value in explaining nature. Data plays the important role of
guiding physicists in the right direction and posing new puzzles that
need to be resolved, keeping the scientific process honest and
exciting.  The disappointment from failures to explain puzzling data
is a crucial aspect of our learning experience, as it encourages
creative individuals to come up with a new way of thinking about the
physical reality.  Over long periods of time, decades or more, the
benefits from a data-driven culture without programmatic reins are so
great that even profit-oriented businesses may choose to support it.

For example, Bell Labs recognized the virtues of such a culture
in the 1930s-70s.  This corporation assembled a collection of creative
scientists in the same corridor, gave them freedom, and harvested some
of the most important discoveries in science and technology of the
20th century, including the foundation of radio astronomy in 1932, the
invention of the transistor in 1947, the development of information
theory in 1948, the solar cells in 1954, the laser in 1958, the first
communications satellite in 1962, the charged-coupled device (CCD) in
1969, and the fiber optic network in 1976. Such long-term benefits
require patience and the foresight of paying it forward.  Investments
in centers of excellence, hosting creative individuals without a
programmatic goal-oriented agenda, establishes a fertile ground for
major breakthroughs.  {\it If the business world recognized the value
of such a culture, shouldn't scientific funding agencies recognize it
as well?}

Christopher Columbus was funded by the Spanish crown to find a new
trade route to the East Indies by sailing westward, but he discovered
the new world of America instead. The funding agency in this case
clearly benefited from his unexpected discovery, as he claimed parts
of America for the Spanish Empire.  Sure, it is important to justify
flagship scientific missions by what we expect to find, but we should
fund them mainly because they might open a new window for unexpected
discoveries.

\section{Opening New Windows of Exploration into the Universe}

In the early 1960s, a panel of ``experts'' was assembled by NASA to
evaluate the merit of a proposal to launch an X-ray telescope into
space. The panel concluded that the scientific justification for such
a mission was weak, since all we could expect to observe are stars
like the Sun emitting in X-rays. The proposal was therefore
rejected. After a decade of delay {\it Uhuru}, the first X-ray
astronomy satellite, was launched. Contrary to expectations from the
original panel of experts, we now know that the X-ray sky is rich and
contains accreting black holes, supernova remnants, galaxy clusters,
and many other unexpected sources. The lesson is simple: whenever
there is a technological opportunity to open a new window for
exploring the Universe, we should open this window without hesitation
since, like Columbus, we might discover new territories that were not
anticipated.

An example for a future window is gravitational-wave astrophysics.
The proposed space-based mission {\it eLISA/NGO}
(http://www.elisa-ngo.org/) expects to find black hole binaries in
galactic nuclei across cosmic time, but it is possible that we will
discover instead new sources that are not being imagined at the moment
and these discoveries will revolutionize physics in the century to
come. Unfortunately, the funding agencies do not share this vision and
{\it eLISA/NGO} has not currently been funded.

In contrast, resources are abundant for projects with predictable
results.  Funding agencies are willing to invest over a billion
dollars on the specific programmatic question: {\it is the energy
density of the vacuum constant over cosmic time to within a percent?}
This programmatic goal is guaranteed to yield results. The problem is
that the range of possible outcomes is defined too
narrowly. Restricted by programmatic reins, large teams of astronomers
are aiming to reduce vast amounts of data with limited attention to
the possibility of unexpected discoveries in aspects of the data that
are not related to their main business agenda. This situation is
analogous to Columbus sailing away from America and ignoring any
unexpected territory which is not the East Indies.

Obviously, agenda-driven projects also lead to important long-term
benefits.  The recent discovery of the Higgs boson in CERN culminated
out of a programmatic experimental effort to confirm a theoretical
idea proposed in the 1960s which lies at the foundation of the
standard model of particle physics.  Although anticipated, this
discovery opens the door to major future advances in unforeseen
directions.  Recognizing the important role of goal-oriented projects,
I am not advocating that funding agencies should shift their primary
focus to open research but rather that they should not ignore it
altogether.\footnote{For example, NASA asks proposers to list the key
milestones they anticipate to accomplish. This request stands in
conflict with the unexpected nature of innovative research, and can
only be respected by proposers who take no risks.} Indeed, Bell Labs
continued to operate as a profit-oriented business during its
innovation period, conservatively manufacturing goods that consumers
buy, so that it could afford to allocate a small fraction of its
revenues towards high-risk research.

\section{Progress is not Linear in Time or Invested Effort}

A few years ago, one of my PhD students worked with me on an elaborate
project that took a year to complete. When the student showed me the
first draft of our paper, I left many comments for him on the
hardcopy. One of my comments was related to the Introduction section
of the paper, in which we described the existing literature on the
subject of our research. My comment said: ``{\it Please add a
reference that discusses a particular possibility that we appear to
ignore in our work}''. The student came back to me a day later and
replied: ``{\it Sorry, but there is no paper in the literature
discussing this novel possibility}''. We immediately realized that
this unexplored idea would be an excellent target for an exciting
follow-up project. We ended up writing a short paper that was
published a few months later in one of the most prestigious journals
for fundamental physics. When the student presented the research at
his PhD research exam, he dedicated most of his talk to the first
project and only a short amount of time at the end to the second
project.  In other words, he chose to organize his discussion based on
the amount of time that it took to complete these two papers, rather
than based on their scientific merit. After his exam, I told him:
``{\it Forget about the long project we worked on for a year. In your
next presentation at a scientific conference, just focus on the
exciting unexpected idea that we came across for our second
project}''.

Progress is not linear in time and sometimes it is even inversely
proportional to the contemporaneous level of invested effort. This is
because progress rests on lengthy preparatory work which lays the
foundation for a potential discovery. Therefore, it is inappropriate
to measure success based on the contemporaneous level of allocated
resources.  Lost resources (time and money) should never be a concern
in a culture that is not tied to a specific programmatic agenda,
because the long-term benefits from finding something different from
what you were seeking could be at an elevated level, far more valuable
than these lost resources. This echoes a quote from 1 Samuel (Chapter
9, 20), concerning the biblical story of Saul seeking his lost
donkeys. The advice Saul received from Samuel, the person who crowned
him as a king after their chance meeting, was simple: ``{\it As for
the donkeys you lost three days ago, do not worry about them...}''.

\bigskip
\bigskip

\acknowledgements I thank M. Dierickx, L. Hernquist, I. Liviatan and
N. Zonnevylle for helpful comments on the manuscript.

\bigskip
\bigskip
\bigskip

\noindent{\underline{\bf Further Reading}}

\begin{itemize}

\item Gertner, J., ``True Innovation'', NY Times Sunday Review,
  February 25, (2012); available online

\item Isaacson, W., ``Inventing the Future'', NY Times Sunday Book
Review, April 6 (2012); available online

\item Loeb, A., ``Taking the Road Not Taken: on the Benefits of
Diversifying Your Academic Portfolio'', Nature {\bf 467}, 358 (2010);
preprint arXiv:1008.1586

\item Loeb, A., ``Rating Growth of Scientific Knowledge and Risk from
Theory Bubbles'', Nature {\bf 484}, 279 (2012); preprint arXiv:1108.5282

\end{itemize}

\end{document}